# An Encryption Scheme with DNA Technology and JPEG Zigzag Coding for Secure Transmission of Images

Grasha Jacob, A. Murugan

**Abstract**—The Internet is a ubiquitous and affordable communications network suited for e-commerce and medical image communications. Security has become a major issue as data communication channels can be intruded by intruders during transmission. Though, different methods have been proposed and used to protect the transmission of data from illegal and unauthorized access, code breakers have come up with various methods to crack them. DNA based Cryptography brings forward a new hope for unbreakable algorithms. This paper outlines an encryption scheme with DNA technology and JPEG Zigzag Coding for Secure Transmission of Images.

—————— ◆ ——————

## 1 INTRODUCTION

INFORMATION Technology plays an important role in convergence of computing, communication and applications to satisfy ever-challenging needs and has influenced and changed every aspect of human lives. Internet provides essential communication between tens of millions of people and is being increasingly used as a tool in the fields of medicine and e-commerce. Today many communication channels are intruded by intruders and ensuring security has become increasingly challenging. As traditional cryptographic methods built upon mathematical and theoretical models are vulnerable to certain attacks, the concept of using DNA computing in the field of cryptography has been identified as a possible technology that brings forward a new hope for unbreakable algorithms.

DNA stands for Deoxyribo Nucleic Acid. DNA is a polymer made of monomers called deoxyribo nucleotides. Each nucleotide consists of three basic items: deoxyribose sugar, phosphate group and a nitrogenous base. The nitrogenous bases are Adenine, Guanine, Cytosine and Thymine. The key thing to note about the structure of DNA is its inherent complementarity proposed by Watson and Crick. A binds with T and G binds to C. All DNA computing applications are based on Watson-Crick complementarity. DNA computing is an interdisciplinary area concerned with the use of DNA molecules for the implementation of computational processes. Adleman's pioneering work [1] gave an idea of solving the directed Hamiltonian Path Problem (Travelling Salesman Problem) of size n in O(n) using DNA molecules. The principle used by Adleman lies in the coding of information (nodes, edges) in DNA clusters and in the use of enzymes for the simulation of simple calculations. The various operations performed on DNA are synthesis, cutting, ligation, translation, substitution, polymerase chain reaction, detection using gel electrophoresis and affinity purification. Gehani et. al., introduced the first trial of DNA based Cryptography in which a substitution method using libraries of distinct one time pads, each of which defines a specific, randomly generated, pair-wise mapping and an XOR scheme utilizing molecular computation and indexed, random key strings were used for encryption[2]. The various work done by [3] - [8] point to the new opportunities of DNA computing in the field of cryptography.

Research work is being done on DNA Computing either using test tubes biologically or simulating the operations of DNA using computers (Pseudo or Virtual DNA computing) [9].

## 2 DNA BASED IMAGE REPRESENTATION

A digital image can be considered a matrix, whose row and column indices identify a point in the image and the corresponding matrix element value identifies the gray level at that point. As there are four bases A, C, T, G in DNA sequence, according to the DNA digital coding

————————————————
- *Grasha Jacob is with the Department. of Computer Science, Rani Anna Government College for Women, Tirunelveli and is a Part-time Research Scholar at the Research and Development Centre, Bharathiar University, Coimbatore – 641046, India. E-mail: grasharanjit@gmail.com*
- *A. Murugan is with the Department of Computer Science, Dr. Ambedkar Government College, Vyasarpadi, Chennai amurugan1972@gmail.com*

technology [10], C denotes the binary value 00, A denotes 01, T denotes 10 and G denotes 11. Therefore any image can be represented in its equivalent DNA form using the DNA digital coding technology.

## 3 JPEG ZIGZAG ENCRYPTION SCHEME

Symmetric encryption algorithms use an identical secret key for encryption and decryption process and the key is sent to the receiver through a secure communication channel.

$$Ke = Kd = K \quad \ldots\ldots\ldots (1)$$

The proposed encryption scheme is a modified version of the hybrid encryption scheme[11] and is a combination of a cryptosystem using JPEG zigzag coding and the DNA based Implementation of YAEA Encryption Algorithm proposed by Amin et. al[12]. The main disadvantage of the hybrid encryption scheme is that the key image used for encryption should be of the same size as that of the image that is to be transmitted. In the proposed encryption scheme using JPEG zigzag coding technology, the

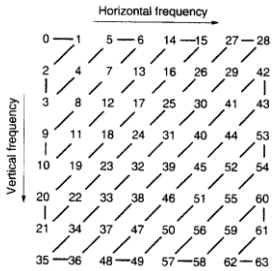

Fig. 1. JPEG Zigzag coding

image to be encrypted is synthesized - transformed into

DNA image and is then translated using JPEG zigzag coding technology as represented in figure 1. The encrypted image is obtained by substituting each quadruple DNA nucleotides sequence of the translated image by one of the many positions of the quadruple nucleotides sequence which is randomly obtained from the gene sequence binary file and the resultant encrypted image is sent to the receiver.

Both the sender and receiver should download the same gene sequence gi| 417839630| ref| NZ_AFQN01000062.1| Haemophilus haemolyticus M19107 M19107_062, whole genome shotgun sequence, 81211 bp linear DNA from GenBank and store the sequence as a binary file. The main advantage of this scheme is that there is no need of the key image and the size of the encrypted image is the same as that of the original image.

The encryption algorithm can be summarized as follows:

**ALGORITHM DNA_JPEG_ZIGZAG_CRYPT**
**Input:** X [image file] to be encrypted, R[Binary file that contains DNA nucleotides sequence]
**Output**: Encrypted image E
1. **SYNTHESIS**
Convert the image X into its corresponding DNA image.
   X ← DNA[X]
2. **TRANSLATION**
   X ← X translated according to JPEG zigzag coding technology
3. **DETECTION and SUBSTITUTION**
For each quadruple DNA nucleotide sequence in X, search starting from a random location in a binary file R represented in the form of a single strand DNA sequence. If the correct pattern is found, its location I is then recorded If search is successful
   Then store I in E;
   Else
   Repeat step 3.
**End Algorithm**

The decryption algorithm can be summarized as follows:

**ALGORITHM DNA_JPEG_ZIGZAG_DECRYPT**
**Input:** Encrypted image E, R[Binary file that contains DNA nucleotides sequence]
**Output:** Decrypted image X
1. **SYNTHESIS**
F ← DNA sequence represented by E[I] in the binary file R
2. **TRANSLATION**
   F ← Reverse JPEG zigzag coding of F
3. **SUBSTITUTION**
Convert F into its binary equivalent and display the decrypted image X.
**End Algorithm**

## 4 EXPERIMENTAL RESULTS

Matlab R2008a is used to simulate the DNA operations on a MiTAC Notebook PC with Intel® Core™ 2Duo CPU T6400 @ 2.00 GHz, 2 GB RAM, 32 bit operating system.
Experiments are performed using different images of different sizes to prove the validity of the proposed algorithm. The experimental results and security analysis show that the proposed algorithm is easy to be implemented, has good encryption effect, strong sensitivity to the DNA sequence file used, and has the abilities of resisting exhaustive, differential and statistical attacks.

## 5 SECURITY ANALYSIS

A good information security system should be able to protect confidential images. The level of security that the proposed encryption algorithm offers is its strength. Cryptographic attacks are a part of cryptanalysis and are designed to subvert the security of cryptographic algorithms, and they are used to attempt to decrypt data without prior access to a key. As a good encryption technique should be robust against statistical, cryptanalytic and differential attacks, the proposed method is examined through these attacks.

## 5.1 STATISTICAL ANALYSIS

The encrypted image should not have any statistical similarity with the original image to prevent the leakage of information. The stability of the proposed method is ex-

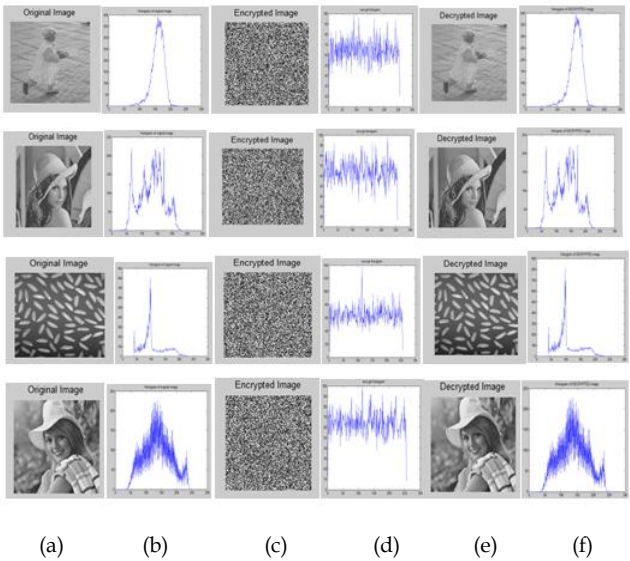

(a)    (b)    (c)    (d)    (e)    (f)

Fig. 2 a) Original Image  b) Histogram of original image  c) Encrypted Image  d) Histogram of encrypted image    e) Decrypted Image    f) Histogram of decrypted image amined via statistical attacks - the histogram and correlation between adjacent pixels.

### 5.1.1  HISTOGRAM ANALYSIS

An image histogram describes how the image-pixels are distributed by plotting the number of pixels at each intensity level. The histograms present the statistical characteristics of an image. If the histograms of the original image and encrypted image are different, then the encryption algorithm has good performance, n intruder cannot extract the pixels' statistical nature of the original image from the encrypted image and the algorithm can resist a chosen plain image or known plain image attack.

Fig 2 reveals that the histograms of the encrypted images are fairly uniform and significantly different from that of the original image. As the encrypted images are almost similar and do not provide any information regarding the distribution of gray values to the intruder, the proposed algorithm can resist any type of histogram based attacks and strengthens the security of the encrypted images significantly.

## 5.1.2 CORRELATION COEFFICIENT ANALYSIS

In most of the plaintext-images, there exists high correlation among adjacent pixels, while there is a little correlation between neighboring pixels in the encrypted image. It is the main task of an efficient image encryption algorithm to eliminate the correlation of pixels. Two highly uncorrelated sequences have approximately zero correlation coefficient and two strongly correlated sequences have a correlation coefficient nearly equal to one.

The Pearson's Correlation Coefficient is determined using the formula:

$$r = \frac{n\sum xy - (\sum x)(\sum y)}{\sqrt{n(\sum x^2) - (\sum x)^2} \sqrt{n(\sum y^2) - (\sum y)^2}}$$

… … …    (2)

where x and y are the gray-scale values of two adjacent pixels in the image and N is the total number of pixels selected from the image for the calculation.

Fig 3 (a) to (g) represent the correlation between the adjacent pixels of the original and encrypted images columnwise, row-wise and diagonal-wise. The correlation of the encrypted images is almost uniformly distributed and gives no information to the intruder regarding the nature of the original image that is being transmitted.

Table 1 tabulates the correlation coefficient calculated for

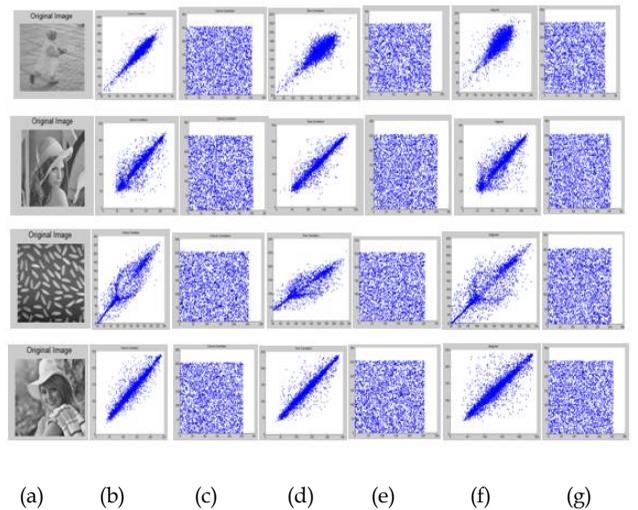

(a)    (b)    (c)    (d)    (e)    (f)    (g)

Fig. 3 a) Original Image  b) Correlation coeff columnwise of original image  c) Correlation coeff columnwise of Encrypted Image   d) Correlation coeff rowwise of original image   c) Correlation coeff  rowwise of Encrypted Image  f) Correlation coeff diagonalwise of original image   g) Correlation coeff diagonalwise of Encrypted Image the original and encrypted images.  It is clear from Table 1 that there is negligible correlation between the two adjacent pixels in the encrypted image and gives no information to the intruder regarding the nature of the original image that is being transmitted. However, the two adja-cent pixels in the original image are highly correlated.

## 5.2 Cryptanalysis

Security is linked to the ability to guess the values of the encrypted data. For example, from a security point of view, it is preferable to encrypt the bits that look the most random.

### 5.2.1 Chosen/Known-plain text attack

For encryption with a higher level of security, the security against both known-plaintext and chosen-plaintext attacks are necessary. Chosen/Known-plain text attacks are such attacks in which one can access/choose a set of plain texts and observe the corresponding encrypted texts.

A mask image, M is obtained by XOR-ing the plain image C with its corresponding encrypted image $C_1$.
Let $Z_1$ be the encrypted image of the plain text image Z.

$$M \leftarrow C \oplus C_1 \qquad \ldots \ldots \ldots \quad (3)$$

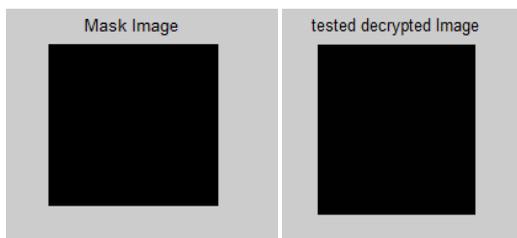

Fig 4 a) XOR Mask   b) Failed attack to crack the Encrypted image

.

```
If Z = M ⊕ Z₁
then
    Unknown encrypted image is decrypted
else
    proposed encryption scheme resists
    Chosen/Known Plain Text attack
        end if                    ……… (4)
```

Fig 4 b) shows an unsuccessful chosen/known-plain text attack using the proposed algorithm.

### 5.2.2 Brute Force Attack

A Brute Force Attack is a strategy that can be used against any encrypted data by an intruder who is unable to take advantage of any weakness in an encryption system that would otherwise make his task easier. It involves systematically checking all possible keys until the correct key is found. The encrypted file is actually randomly generated pointers to the dna sequence file and the possibility of more than one quadruple nucleotide sequence pointing to the same position in the dna sequence file is less. Moreover, the aspect of bio-molecular environment is more difficult to access as it is extremely to recover the DNA digital code without knowing the correct coding technology used. An incorrect coding will cause biological pollution, which would lead to a corrupted image.

TABLE 1
CORRELATION COEFFICIENTS

| File Name | | Johnson.bmp | Lena.bmp | Rice.bmp | Elaine.bmp |
|---|---|---|---|---|---|
| $\gamma_{col}$ | original image | 0.92598 | 0.88478 | 0.9288 | 0.92835 |
| | encrypted image | -0.01552 | 0.00308 | -0.00603 | 0.01652 |
| $\gamma_{row}$ | original image | 0.81756 | 0.94345 | 0.89414 | 0.94625 |
| | encrypted image | 0.028883 | -0.01635 | -0.00692 | -0.006182 |
| $\gamma_{dia}$ | original image | 0.79782 | 0.86315 | 0.87951 | 0.87296 |
| | encrypted image | -0.00526 | 0.00914 | 0.01567 | 0.00783 |

## 5.3 Differential Attack

The aim of differential analysis is to determine the sensitivity of encryption algorithm to slight changes. If an attack is made to create a small change in the plain image to observe the results, this manipulation should cause a significant change in the encrypted image and the opponent should not be able to find a meaningful relationship between the original and encrypted image with respect to diffusion and confusion. A different sequence used or a small change made to the DNA sequence file or the original plain image will result in a completely different encrypted image showing that the algorithm is highly sensitive to slight changes.

## 6 CONCLUSION

In today's corporate world, images travel widely and rapidly, in multiple manifestations, through email and across the Internet. Telemedicine has become a common method for the transmission of images and patient data across long distances. In business transactions, sensible data such as pin numbers are sent as images. Corporations have very little visibility into exactly where their documents are being accessed or by whom. Protecting sensitive information is an ethical and legal requirement.

DNA based encryption helps in the secure transmission of such confidential images. The proposed encryption scheme with DNA technology using JPEG zigzag coding is easy to implement and can resist brute-force, statistical and differential attack and therefore is suitable for multi-level security applications of today's network.

**Grasha Jacob** received her MCA Degree from Avinashilingam University, Coimbatore, India in 1994 and M.Phil Degree in Computer Science from Manonmaniam Sundaranar University in 2003. She is working as an Associate Professor of Computer Science in Rani Anna Government College, Tirunelveli, India. Her research interests include Information Security, Image Processing, Molecular Computing.

**A.Murugan** received his MSc Degree (Gold Medalist) from Manonmaniam Sundaranar University, Tirunelveli, India in 1994 and Ph.D from University of Madras, Chennai, India in 2005. He is working as Associate Proessor in the Department of Computer Science, Dr. Ambedkar Government College, Vyasarpadi, Chennai, India. He has published two books. He has published four papers in the International Journal of Computer Mathematics. His research interests include Molecular Computation, Graph Theory, Data Structure, Analysis of Algorithms and Theoretical Computer Science He is a member of the Editorial Board of International Journal of Advanced Computer Science and Technology (IJACST) and International Journal of Statistics and Analysis (IJSA) – Research India Publications.